\documentclass[10pt, conference, letterpaper]{IEEEtran}
\IEEEoverridecommandlockouts
\usepackage{amsmath,amssymb,amsfonts}
\usepackage{algorithmic}
\usepackage{graphicx}
\usepackage{textcomp}
\usepackage{xcolor}
\usepackage{multirow}
\usepackage{booktabs}
\usepackage{subfig}
\usepackage[ruled,vlined,linesnumbered]{algorithm2e}
\usepackage[numbers,sort&compress]{natbib}
\usepackage{hyperref}
\def\BibTeX{{\rm B\kern-.05em{\sc i\kern-.025em b}\kern-.08em
    T\kern-.1667em\lower.7ex\hbox{E}\kern-.125emX}}
    
\SetKwInput{KwInput}{Input}                
\SetKwInput{KwOutput}{Output}              
\SetKwComment{Comment}{$\triangleright$\ }{}

\begin{document}

\title{Dual Attention-Based Federated Learning for Wireless Traffic Prediction
}

\author{\IEEEauthorblockN{Chuanting Zhang, Shuping Dang, Basem Shihada, Mohamed-Slim Alouini}
\IEEEauthorblockA{
Computer, Electrical and Mathematical Science and Engineering Division, \\
King Abdullah University of Science and Technology\\
Thuwal, Saudi Arabia \\
Email: \{chuanting.zhang, shuping.dang, basem.shihada, slim.alouini\}@kaust.edu.sa}
}

\maketitle

\begin{abstract}
Wireless traffic prediction is essential for cellular networks to realize intelligent network operations, such as load-aware resource management and predictive control.
Existing prediction approaches usually adopt centralized training architectures and require the transferring of huge amounts of traffic data, which may raise delay and privacy concerns for certain scenarios.
In this work, we propose a novel wireless traffic prediction framework named \textit{Dual Attention-Based Federated Learning} (FedDA), by which a high-quality prediction model is trained collaboratively by multiple edge clients.
To simultaneously capture the various wireless traffic patterns and keep raw data locally, FedDA first groups the clients into different clusters by using a small augmentation dataset.
Then, a quasi-global model is trained and shared among clients as prior knowledge, aiming to solve the statistical heterogeneity challenge confronted with federated learning.
To construct the global model, a dual attention scheme is further proposed by aggregating the intra- and inter-cluster models, instead of simply averaging the weights of local models.
We conduct extensive experiments on two real-world wireless traffic datasets and results show that FedDA outperforms state-of-the-art methods.
The average mean squared error performance gains on the two datasets are up to 10\% and 30\%, respectively.
\end{abstract}


\section{Introduction}
Since the commercialization of fifth generation (5G) communication networks in 2019, the preliminary research on the potential features and enabling technologies for the sixth generation (6G) communications has attracted extensive attention in academia and industry \cite{David2018, Zong2019}. There are a set of emerging technologies and novel paradigms, e.g., terahertz spectrum, space-air-ground communications, large reflecting surfaces, and cognitive radios \cite{dang2020should}. In addition, the communication research community has nevertheless reached a consensus that artificial intelligence (AI) is the key to implement novel paradigms, coordinate heterogeneous networks, organize various communication resources, and enable the truly smart 6G communications in the 2030s \cite{Letaief2019}. In particular, AI aided by big data and high-rate real-time transmission capability is expected to be the most efficient approach to reduce network overhead and elevate the quality of service (QoS) of both access and core networks \cite{Yao2019}. 

To facilitate the fusion of AI and communication networks, wireless traffic prediction is indispensable. 
Wireless traffic prediction \cite{Wang2017, Zhang2019} estimates traffic data volume in the future and provides the decision basis of communication network management and optimization \cite{Xu2019}. With the predicted traffic data, proactive measures can be taken to mitigate the network congestion and outage caused by burst transmissions. 
Moreover, the heterogeneous service requirements, which are expected to become common in 6G communication networks \cite{Kato2020}, can be well satisfied with a lower cost by wireless traffic prediction. This will lead to a significant improvement in the QoS from both network's and user's perspectives. 

Currently, most of wireless traffic prediction approaches are focusing on centralized learning strategy and involves transferring huge amount of raw data to a datacenter to learn a generalized prediction model. However, frequently transmission of training data and signaling overhead could easily exhaust the network capacity and yield negative impacts on payload transmissions. Thus new wireless traffic prediction approaches that can cope with the above challenges are needed. 

The emergence and success of federated learning (FL) \cite{McMahan2017, yang2019federated, li2019federated, Bonawitz2019, kairouz2019advances} make the prediction problem possible while keep data locally. In FL setting, many clients e.g., mobile devices, base stations (BSs), or companies, collaboratively train a prediction model under the orchestration of a central server. Only intermediate gradients or model parameters obtained by local training are sent to the central server, instead of the raw data. There are enough reasons to support FL in the next-generation communications \cite{Tran2019a}. First, the advances of edge computing have paved the way for easily implementing FL in reality. As edge clients equip abundant computing resources \cite{Liu2018a}, the centralized powerful datacenter is no longer essential and the delay of transferring raw data can be considerably reduced. In addition, FL facilitates unprecedented large-scale flexible data collection and model training. The edge clients can proactively collect data during day hours, then jointly update the global model during night hours, to improve efficiency and accuracy for next-day usage.

Despite the promising application prospect, accurate wireless traffic prediction under the FL settings is still a major research challenge, especially the network-wide prediction. 
This is because user mobility can cause sophisticated spatio-temporal coupling among wireless traffic, which can hardly be captured and modeled. 
Furthermore, different BSs may have distinct traffic patterns which makes the traffic data highly heterogeneous and learning and prediction on this kind of heterogeneous data is very challenging.

Therefore, to cope with the wireless traffic prediction issues for future communication networks, we propose a novel wireless traffic prediction framework named the dual attention-based federated learning (FedDA), by which a high-quality prediction model is trained collaboratively by multiple BSs. The FedDA framework relies on a set of state-of-the-art training paradigms, including a data augmentation assisted clustering strategy, an intermediate and auxiliary training model, a dual attention-based model aggregation, and a hierarchical aggregation structure. 
Specifically, the processing of FedDA can be split into three stages to ensure high-accuracy, transferable, and secure training process for wireless traffic prediction. 

We first introduce an augmentation assisted clustering strategy to group all BSs, i.e., clients in the context of FL, into a number of clusters depending on their augmented traffic patterns and geographic locations. Then, leveraging the augmented data collected from distributed BSs, a quasi-global prediction model can be constructed at the central server. This quasi-global model is used to mitigate the generalization difficulty of the global model caused by the statistical heterogeneity among traffic data collected from different clusters. Finally, instead of simply averaging the model weights collected from local clients to yield the global model, a dual attention-based model aggregation mechanism and a hierarchical aggregation structure are adopted at the central server. By introducing the dual attention and hierarchical settings, an adequate equilibrium between generality and specialty can be achieved.

Following the proposed FedDA framework and the descriptions given above, the contributions of this paper can be summarized as follows:

\begin{itemize}
\item We design a data augmentation-assisted iterative clustering strategy, which takes the augmented data and geographic locations of clients as clustering reference to simultaneously capture various traffic patterns of clients and protect data privacy.
\item We introduce a quasi-global model, which is an intermediate and auxiliary tool to mitigate the generalization difficulty of the global model caused by the statistical heterogeneity among traffic patterns collected from different clients.
\item We propose the FedDA framework consisting of two advanced settings for aggregations, which are the dual attention-based model aggregation mechanism and the hierarchical aggregation structure. In this way, the central server can capture not only the cluster-specific data patterns but also ensure the transferability of the global model.
\item We verify the effectiveness and efficiency of the FedDA framework by testing on two real-world datasets and compare the experimental results with those generated by existing algorithms.
\end{itemize}

The rest of this paper is organized as follows. Section II reviews related works on wireless traffic prediction and FL. Section III gives the system model and problem statement. In section IV, we introduce our proposed dual attention mechanism in detail, including the data augmentation-assisted client clustering, mathematical expression of dual attention, and the corresponding optimization techniques. Section V presents and discusses all the experiments. Finally, Section VI concludes this paper. Our code is available at \url{https://github.com/chuanting/FedDA}.

\section{Related Work}
As the present work is closely related to wireless traffic prediction and FL, we review the most related achievements and milestones of these two research topics in this section.
\subsection{Wireless Traffic Prediction}
Recently, wireless traffic prediction has received a lot of attention as many tasks in wireless communications require accurate traffic modeling and prediction capabilities.
Wireless traffic prediction is essentially a time series prediction problem. The methods to solve it can be roughly classified into three categories, i.e., simple methods, parametric methods, and non-parametric methods.

Historical average and na\"ive methods are representatives of the first category \cite{hyndman2018forecasting}. The former predicts all future values as the average of the historical data, while the latter takes the last observation as the future. This kind of prediction method involves no complex computations, and thus makes it quite simple and easy to implement. However, as simple methods fail to capture the hidden patterns of wireless traffic, their prediction performance is relatively poor.

For the second category, i.e., parametric methods, the wireless traffic is modeled and predicted based on tools from statistics and probability theory. The most classical method is AutoRegressive Integrated Moving Average (ARIMA) \cite{hamilton1994time}.
To characterize the self-similarity and bursty of wireless traffic, ARIMA and its variants were explored in \cite{Shu2005, zhou_ngide_2006}.
In a recent study, \cite{Xu2016} first decomposed the wireless traffic into regularity and randomness components. Then the authors demonstrated that the regularity component can be predicted through the ARIMA model, but the prediction of random components is impossible.
Besides the ARIMA model, the $\alpha$-stable model \cite{Li2017}, entropy theory \cite{Li2014}, and covariance functions \cite{Chen2015} were also explored to perform wireless traffic prediction.

As machine learning and AI techniques \cite{Liu2018} continue their fast evolving, the non-parametric methods have made themselves strong competitors to parametric methods for wireless traffic prediction. 
Particularly, recent years have witnessed an obvious trend in solving wireless traffic prediction problems based on deep neural networks.
In \cite{Nie2017}, the authors proposed a deep belief networks-based prediction method for wireless mesh networks. In \cite{Wang2017}, A hybrid deep learning framework was designed on the basis of autoencoder and Long-Short Term Memory networks (LSTM) to simultaneously capture the spatial and temporal dependence among different cells.
Aiming to perform prediction on multiple cells, the researcher also introduced a multi-task learning framework by using LSTM \cite{Qiu2018}.
Besides, the city-scale wireless traffic predictions are also investigated in \cite{Zhang2018, Zhang2019}, in which the authors introduced novel prediction frameworks by modeling spatio-temporal dependence over cross-domain datasets.

All aforementioned works mainly focus on wireless traffic prediction in the centralized way. Our proposed framework in this paper differs from the above works, and we are trying to solve the wireless traffic prediction problem by a distributed architecture and federated learning.

\subsection{Federated Learning}

FL provides a distributed training architecture that can be jointly applied with many machine learning algorithms, in particular deep neural networks.
In FL, a global model can be obtained by aggregating local clients' models. To obtain the global model, \cite{McMahan2017} introduced an aggregation method called federated averaging (FedAvg). 
Research shows that when the client data is independent and identically distributed (IID), FedAvg achieves similar performance compared with centralized learning. However, when the client data is non-IID, the performance of FedAvg degrades greatly. 
To solve this problem, \cite{zhao2018federated} proposed a data-sharing strategy by creating a small subset of data which is globally shared among all the client devices. This strategy can solve the statistical heterogeneity challenge confronted with FL.
In \cite{Li2020}, the authors introduced FedProx, which can be viewed as a generalization and re-parameterization of FedAvg, to tackle heterogeneity in federated networks.
Besides, \cite{Ji2019} introduced an attentive federated aggregation scheme, called FedAtt, by considering unequal contributions from different clients to the global model. This scheme improves the generalization ability of the global model and has been successfully used to solve the language modeling problem.
For a more detailed introduction on the development of FL, please refer to a recent survey \cite{li2019federated}.

Our work is inspired by the above research, but we mainly focus on wireless traffic prediction problem, which is different from the above works. Also,  
because the wireless traffic data is highly heterogeneous, we propose a novel FedDA framework to solve this statistical challenge.


\section{Problem Formulation and Preliminaries}
In this section we provide the problem formulation of wireless traffic prediction and give the implementation details of FL corresponding to the formulated problem.

\subsection{Problem Formulation}
Given $K$ BSs, each BS have its own local wireless traffic data, denoted as $d^k=\{d^k_1,d^k_2,\cdots,d^k_Z\}$ with a total of $Z$ time intervals.
The wireless traffic prediction problem can be described as the prediction of future traffic volume based on the current and the previous traffic volumes.
Suppose $d^k_z$ to be the target traffic volume required to be predicted, then the wireless traffic prediction problem can be described as
\begin{equation}\label{p1}
\hat{d}^k_{z} = f(d^k_{z-1}, d^k_{z-2},\cdots, d^k_1; w),
\end{equation}
where $f(\cdot)$ denote the chosen prediction model and $w$ the corresponding parameters.
The prediction model $f(\cdot)$ can be either in a linear form like linear regression or in a nonlinear form like deep neural networks.

For machine learning based wireless traffic prediction techniques, it is common to use only part of the historical traffic data as input features to reduce complexity.
Thus, based upon $d^k$, a set of input-output pairs $\{x^k_i, y^k_i\}_{i=1}^n$ can be obtained by using sliding window scheme.
$x^k_i$ denotes the historical traffic data related to $y^k_i$, and we set it to $\{d^k_{z-1},\cdots,d^k_{z-p},d^k_{z-\phi 1}, \cdots, d^k_{z- \phi q}\}$.
$p$ and $q$ are the sliding window sizes for capturing the closeness dependence and period dependence of wireless traffic data and $\phi$ is the periodicity \cite{Zhang2017, Zhang2018}.
As we focus only on the one-step ahead prediction problem, so $y^k_i=d^k_z$.
Thus, the problem formulated in (\ref{p1}) can be reformulated as 
\begin{equation}\label{p2}
\hat{y}^k_i = f(x^k_i; w).
\end{equation}
Our objective is to minimize the prediction error over all $K$ BSs, thus the parameters $w$ can be obtained by solving
\begin{equation}\label{p3}
\arg\min_{w}\left\lbrace\frac{1}{Kn}\sum_{k=1}^K\sum_{i=1}^n\mathcal{L} (f(x^k_i; w), y^k_i)\right\rbrace,
\end{equation}
where $\mathcal{L}$ is the loss function and the structure typically takes $|f(x^k_i; w) - y^k_i|^2$ or $|f(x^k_i; w) - y^k_i|$.

\begin{figure}[!t]
\centering
\includegraphics[width=0.35\textwidth]{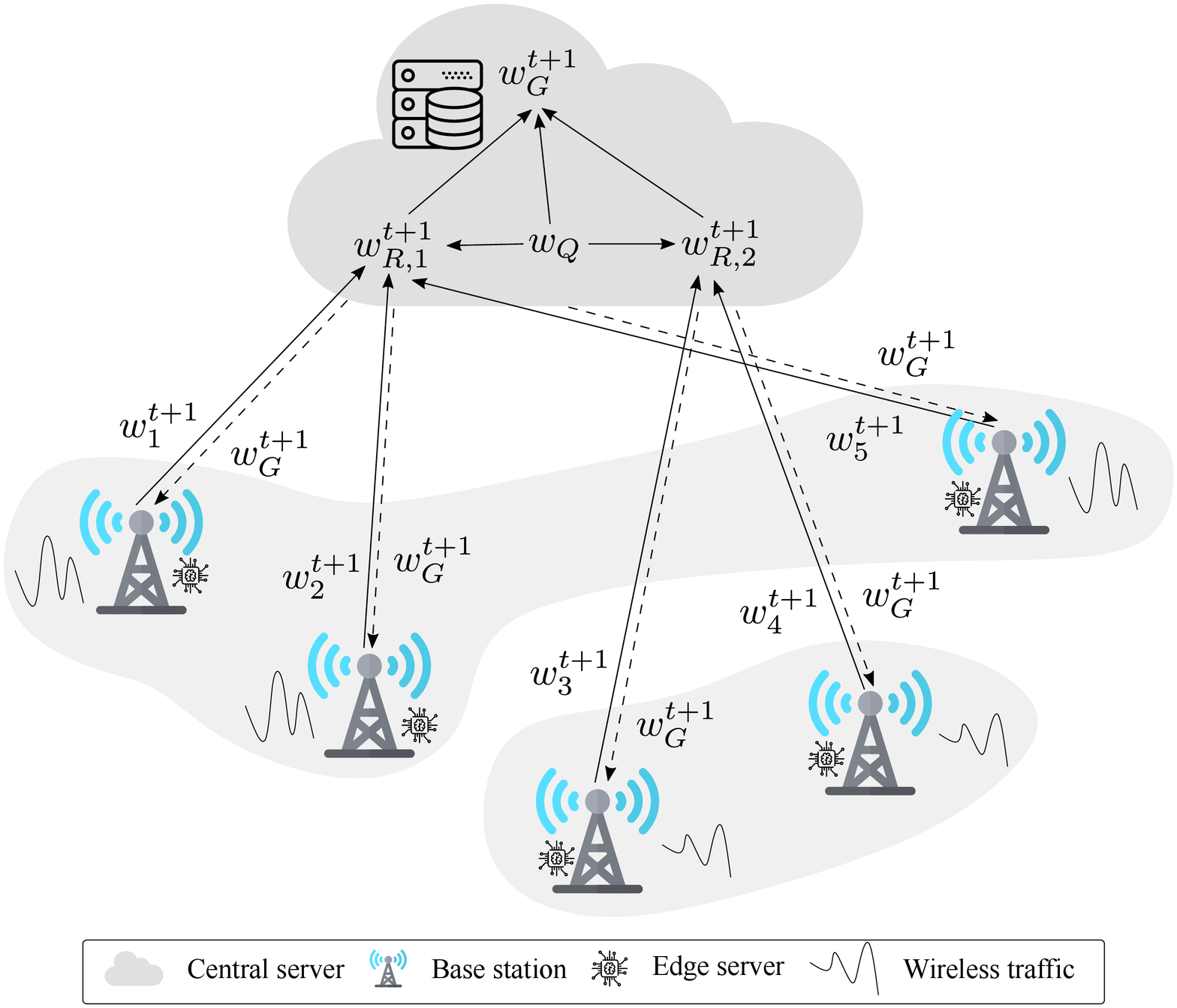}
\caption{System diagram of FedDA with five BSs in two clusters.}
\label{sys_model}
\end{figure}

\subsection{Preliminaries of Federated Learning}
We try to solve (\ref{p3}) in a distributed manner, particularly under the cross-silo FL settings \cite{kairouz2019advances} and assume data are located at geo-distributed clients.
After initializing the global model characterized by the parameters $w$, FL works as follows at the $t$-th training round:
\begin{enumerate}
    \item The central server send the global model $w^t$ to all BSs;
    \item The BS treats the global model as prior knowledge and updates its local model based upon $w^t$ with its local data. The local model update rule is given as follows: $w_k^{t+1} \leftarrow w_k^t - \eta \nabla_{w^t} \mathcal{L}(f(x^k; w^t), y^k)$, where $\eta$ is learning rate and $\nabla_{w^t}$ is the gradient of loss function with respect to $w^t$;
    \item The BS sends $w_k^{t+1}$ to the central server;
    \item The central server performs model aggregation (also known as federated optimization) based on local models. 
    The most classical model aggregation scheme is the federated averaging \cite{McMahan2017}, which can be written as $w^{t+1} \leftarrow \frac{1}{K}\sum_{k=1}^K w_k^{t+1}$.
\end{enumerate}
By running the above steps iteratively until the termination conditions are satisfied, a final global model $w$ can be obtained.

\section{Proposed Framework}
In this section, we give a detailed introduction of our proposed wireless traffic prediction framework: FedDA, and a demo FedDA system diagram with five BSs in two clusters is shown in Fig. \ref{sys_model}. Specifically, FedDA consists of three steps: 
\begin{enumerate}
    \item Each BS performs the traffic augmentation procedure and sends the augmented data to the central server;
    \item The central server yields a quasi-global model and clusters the BSs into different groups based on the augmented data and the location information of BSs;
    \item The BSs cooperatively train a global model under the orchestration of the central server by using the dual attention-based federated optimization.
\end{enumerate}
In the following, we explain how to augment wireless traffic data and analyze the similarities between augmented data and original traffic data.
After that, we introduce the iterative clustering strategy by taking into consideration both locations and traffic patterns of BSs, followed by a detailed elaboration of our proposed dual attention-based model aggregation scheme.

\subsection{Wireless Traffic Data Augmentation}
As urban areas have different functions to support the daily operation of a city, the traffic patterns of spatially distributed BSs differ a lot.
Besides, users have different mobility and communication behaviors, which further enlarge the pattern diversity of wireless traffic.
\begin{figure}[!t]
\centering
\subfloat[Wireless traffic augmentation strategy.]{\includegraphics[width=0.35\textwidth]{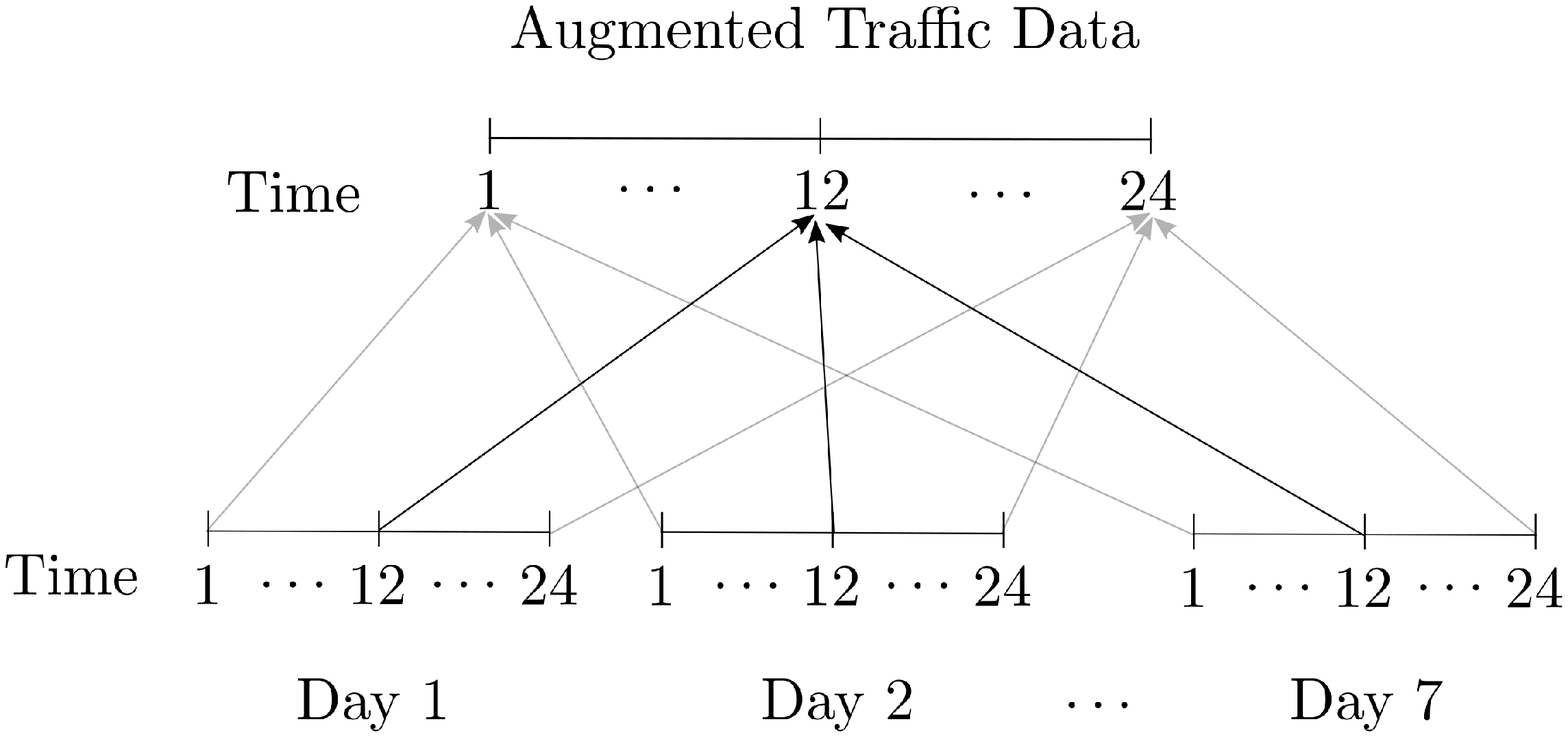}\label{data_augmentation}}

\subfloat[Augmented versus original traffic data.]{\includegraphics[width=0.35\textwidth]{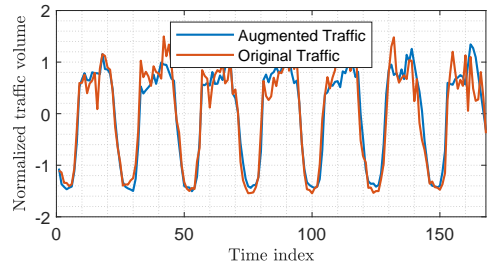}\label{data_vs_raw}}

\caption{Illustration on wireless traffic data augmentation and the comparison between the augmented and the original traffic data of a BS.}
\label{aug_compare}
\end{figure}
Therefore, wireless traffic data of different BSs has a high level of heterogeneity and is essentially non-IID.
Performing FL over non-IID data is quite challenging as weight divergence exists when performing model aggregation on the server side \cite{zhao2018federated}.
Herein, we propose a data sharing strategy by creating a small augmentation dataset using original wireless traffic dataset to conquer the statistical heterogeneity challenge.

Our augmentation strategy works as follows. We first split the dataset into weekly slices based on the time index. For the traffic of each week we compute the statistical average value for each time point and treat the obtained result as the augmented data. Finally, the augmented traffic is standardized to have zero mean and unit variance.
An illustration of the augmentation procedure is displayed in Fig. \ref{data_augmentation}, and the comparison between the augmented wireless traffic and the original traffic can be found in Fig.  \ref{data_vs_raw}.

It can be seen from Fig. \ref{aug_compare} that the proposed augmentation strategy is quite easy to implement and produce the augmented data, compared with traditional time series data augmentation strategies either in time domain or frequency domain \cite{wen2020time}. Though the strategy is simple and straightforward, it works well and achieves a Pearson correlation coefficient of $0.9526$, which indicates high similarities between augmented data and original data.

Each BS sends a very small part, say $\varphi \%$, of its augmented data to the central server, in which a global dataset that consists of a uniform distribution over $y$ is centralized. Note that compared with the size of the raw data, the augmented data size is much smaller. 
Based upon this augmentation dataset, a quasi-global model can be trained and treated as prior knowledge for all BSs.
We use the term `quasi-global' because the model is trained using augmented data of all BSs, instead of the original data.
Even so, this model can still be used as prior knowledge of all BSs because of the high similarities between the augmented data and the original data. 
We characterize the quasi-global model by $w_Q$ in Fig. \ref{sys_model}, and it has exactly the same network architecture as the local models and the global model.

\begin{algorithm}[!t]
\DontPrintSemicolon
\KwInput{BS location $\{g_k\}_{k=1}^K$, augmented traffic $\{\tilde{d}^k\}_{k=1}^K$, cluster size $C$, cluster center $\{v_c\}_{c=1}^C$, and iteration threshold $J$.}
\KwOutput{$C$ clusters}
Random initialize $\{v_c\}_{c=1}^C$\\
  
  \While{$j < J$}
  {
     
     Group $\{\tilde{d}^k\}_{k=1}^K$ into $C$ clusters ($l_1$, $l_2$, $\cdots$, $l_C$) by using K-Means with $\{v_c\}_{c=1}^C$; \\
     Group $\{g_k\}_{k=1}^K$ into $C$ clusters ($l'_1$, $l'_2$, $\cdots$, $l'_C$) by using K-Means \\
     \If { $\{l_c\}_{c=1}^C$ is the same as $\{l'_c\}_{c=1}^C$}
	 {
	 		break
	 }
	 Update cluster center $\{v_c\}_{c=1}^C$ based on ($l'_1$, $l'_2$, $\cdots$, $l'_C$)\\
	 $j = j + 1$
	 
  }
\caption{Iterative clustering strategy}
\label{ics}
\end{algorithm}

\subsection{Iterative Clustering for BSs}
As mentioned earlier, spatially distributed BSs have different traffic patterns.
To capture the pattern heterogeneity among BSs and train an accurate prediction model suited for most BSs, we perform clustering analysis for BSs and propose an iterative clustering strategy to achieve this purpose by taking into consideration both the geo-locations and the traffic patterns of BSs.
The detail of our clustering strategy is summarized in Algorithm \ref{ics}.

For an arbitrary BS $k$, the central server knows its location information $g_k$ (i.e., the longitude and latitude) and stores its augmented traffic data $\tilde{d}^k$. 
By using a random initialization of $C$ cluster centers $\{v_c\}_{c=1}^C$, we perform the K-Means algorithm on the augmented data $\{\tilde{d}^k\}_{k=1}^K$ and obtain the cluster labels of BSs (Line 3 of Algorithm \ref{ics}).
Then we use the location information $\{g_k\}_{k=1}^K$ as input and similarly perform the K-Means algorithm on it. This can yield $C$ different clusters (Line 4 of Algorithm \ref{ics}).
If the clustering results on these two kinds of data are the same, then Algorithm \ref{ics} stops and returns the cluster label of each BS.
If the yielded results are not the same, then based on the obtained cluster label on geo-location data, we compute the cluster center and use this to initialize the K-Means clustering on the traffic data.
This indicates that the geo-location information is considered by the traffic pattern clustering process.
The above steps are repeated on an iterative basis until the termination conditions are satisfied.

As shown in Fig. \ref{sys_model}, the BSs are clustered into different clusters based on their geo-location and traffic pattern.
After obtaining the cluster label of each BS, FedDA proceeds to the federated optimization, which will be introduced in the next subsection.

\subsection{Dual Attention-Based Model Aggregation}
\begin{figure}[!htbp]
\centering
\includegraphics[width=0.32\textwidth]{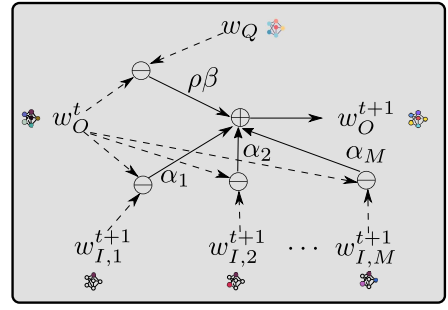}
\caption{Dual attention-based model aggregation.}
\label{fed_dual}
\end{figure}

One of the most fundamental part of FL is the model aggregation scheme on the central server side, which involves constructing the final global model based on the received local ones.
In this paper, we propose a novel federated optimization strategy, i.e., FedDA, for obtaining the global model.
Specifically, we introduce the attention scheme into model aggregation in FedDA and quantifies the contributions of both local models and the quasi-global model in a layer-wise manner.
Fig. \ref{fed_dual} shows an illustration of our proposed layer-wise dual attention-based model aggregation procedure.
Note that we adopt a hierarchical learning scheme in FedDA. That is, there are two levels of model aggregation. The first one performs intra-cluster model aggregation, whose function is to obtain cluster models that capture the unique traffic patterns of each cluster.
The second one performs inter-cluster model aggregation, after which the final global model that incorporates knowledge of all clusters is generated.

In Fig. \ref{fed_dual}, $w_{I,m}^{t+1}$ denotes the $m$-th input model, and there are $M$ input models in total. $w_O^{t+1}$ denotes the output model.
For intra-cluster model aggregation in cluster $c$, $w_{I,m}^{t+1}$ belongs to a local model of that cluster, and $w_O^{t+1}$ stands for $w_{R,c}^{t+1}$.
For inter-cluster model aggregation, $w_{I,m}^{t+1}$ belongs to $\{w_{R,c}^{t+1}\}_{c=1}^C$, and the output is the global model $w_G^{t+1}$. 


The purpose of federated optimization on the central server side is to find an optimal global model that can have a strong generalization ability over all BSs.
To achieve this, the global model should find a balance between capturing the unique and the common-shared traffic patterns of BSs.
Thus, in our proposed scheme, we regard the optimization problem as finding a global model that is close to both of the local models and the quasi-global model in the parameter space, considering their different contributions during the model aggregation procedure.
Consequently, the optimization objective is to minimize the sum-weighted distance among different models by using self-adaptive scores as weights.
The federated optimization problem is formally defined as
\begin{equation}\label{obj}
\arg\min\limits_{w_O^{t+1}}\left\lbrace\sum\limits_{m=1}^M \frac{1}{2}\mathbf{\alpha}_m \mathcal{L}(w_O^t, w_{I,m}^{t+1})^2 + \frac{1}{2}\rho \beta \mathcal{L}(w_O^t, w_Q)^2\right\rbrace,
\end{equation}
where $\alpha_m$ and $\beta$ represent attention weight vectors denoting the contributions of each layer of the $m$-th input model and the quasi-global model; $\rho$ is a task-dependent regularization parameter and can be manually set based on experiment requirements.

To obtain the weights $\alpha_m$, we use the attention mechanism and apply it to the layer-wise parameters.
For the $m$-th input model, the parameter of the $l$-th layer is denoted as $w_{I,m}^l$. Similarly, the parameter of the $l$-th layer of the output model is denoted as $w_{O}^l$. The time stamps are omitted in $w_{I,m}^l$ and $w_{O}^l$ for simplicity.
Based on the layer-wise parameters, the distance between $w_{I,m}^l$ and $w_{O}^l$ can be calculated by the Frobenius norm of their difference, which is expressed as
\begin{equation}\label{distance}
\begin{split}
s_m^l=\mathcal{L}(w_{I,m}^l, w_O^l)= \lVert w_{I,m}^l-w_O^l \rVert_2^2.
\end{split}
\end{equation}

\begin{algorithm}[!t]
\DontPrintSemicolon
\KwInput{Wireless traffic data $\{x^k, y^k\}_{k=1}^K$; Quasi-global model, $w_Q$; Fraction of BSs, $\delta$; Learning rate of local BS, $\eta$; Step size of server side, $\gamma$.}
\KwOutput{Global model, $w_G$}

\For {each round $t=1, 2, \cdots,$ }
		{
			$\mathfrak{m} \leftarrow max (K\cdot \delta, 1)$ \\
			$S_t \leftarrow$ a random set of $\mathfrak{m}$ BSs\\
			\For {each client $k \in S_t$}
			{
				$w_k^{t+1} \leftarrow w_k^t - \eta \nabla_{w^t} \mathcal{L}(f(x^k; w^t), y^k) $
			}
			\For {cluster $c=1,2,\cdots$} 
			{
				$S_c \leftarrow$ a set of BS with cluster label $c$\\
				Obtain $w_{R,c}^{t+1}$ by using Equations (\ref{distance}) to (\ref{global_update})  
			}
			Obtain $w_G^{t+1}$ by using Equations (\ref{distance}) to (\ref{global_update})
		}
\caption{Implementation of FedDA}
\label{fedda_alg}
\end{algorithm}

Subsequently, the softmax function is applied to $s_m^l$ and maps the non-normalized distance values to a probability distribution over the $M$ input models. In this way, the contributions of these models can be determined. The standard softmax function $\sigma(\cdot)$ is described as
\begin{equation}\label{weight}
\begin{split}
\alpha_m^l = \sigma(s_m^l)= \frac{e^{s_m^l}}{\sum_{m=1}^M e^{s_m^l}}.
\end{split}
\end{equation}

Similarly, the values of $\beta$ can be obtained. After we get the $\alpha_m$ and $\beta$, the parameters of the output model can be updated by the gradient descent algorithm. 
We first calculate the derivative of (\ref{obj}) with respect to $w_O^t$ and obtain the corresponding gradient
\begin{equation}\label{gradient}
\nabla = \sum_{m=1}^m\alpha_m (w_O^t - w_{I,m}^{t+1}) + \rho \beta (w_O^t - w_Q).
\end{equation}
With the derived gradient, the output model parameters can be updated by
\begin{equation}\label{global_update}
w_{O}^{t+1} = w_O^t -\gamma (\sum_{m=1}^M\alpha_m (w_O^t - w_{I,m}^{t+1}) + \rho \beta (w_O^t - w_Q)),
\end{equation}
where $\gamma$ is a predetermined step size that controls how much $w_O$ should move in the direction of the opposite gradient in every iteration. The whole procedure of our proposed FedDA is summarized in Algorithm \ref{fedda_alg}.

\begin{table*}[]
\centering
\renewcommand{\arraystretch}{1.3}
\caption{Prediction performance comparisons among different methods in terms of MSE and MAE on two datasets (`$\uparrow$' denotes the performance gain of FedDA over FedAtt).}
\label{tab:performance}
\resizebox{1.0\textwidth}{!}{%
\begin{tabular}{|l|ccc|ccc|ccc|ccc|}
\hline
\multirow{3}{*}{Methods} & \multicolumn{6}{c|}{Milano}                         & \multicolumn{6}{c|}{Trento}                         \\ \cline{2-13} 
                         & \multicolumn{3}{c|}{MSE} & \multicolumn{3}{c|}{MAE} & \multicolumn{3}{c|}{MSE} & \multicolumn{3}{c|}{MAE} \\ \cline{2-13} 
                         & SMS  & Call  & Internet  & SMS  & Call  & Internet  & SMS  & Call  & Internet  & SMS  & Call  & Internet  \\ \hline
Lasso                                           &  0.7580    &  0.3003     &          0.4380 &  0.6231    &   0.4684    &   0.5475        &  4.7363    &   1.6277    &   5.9121 &   1.3182   &   0.8258    &  1.5391         \\ 
SVR                                            &  0.4144    &   0.0919    &          \textit{0.1036} &  0.3528    &  0.1852     &    \textit{0.2220}       &  5.2285    &  1.7919     &  5.9080 & 1.0390     &  0.5656     &  1.0470         \\ 
LSTM                                           & 0.5608 &  0.1379  &          0.1697 & 0.4287   &  0.2458     &    0.2936   &   3.6947    &   1.1378    &    4.6976            &   0.9426   &   0.5013    &    1.1193       \\ 
FedAvg                                         &  0.3744    &  0.0776     &          0.1096 &  0.3386    &   0.1838    &   0.2319        &   2.2287   &   1.6048    &   4.7988        &  0.7416    &   0.5319    &    1.0668       \\ 
FedAtt                                         &  0.3667    &   0.0774    &          0.1096 &  0.3375    &   \textit{0.1837}    &   0.2321        &  2.1558    &   1.5967    &    4.7645       &  0.7444    &  0.5306     &  1.0629         \\ \hline
FedDA ($\varphi$=$1$)                                    &  0.3559    & \textit{0.0752}      &  0.1118         & 0.3353 &  0.1820     &  0.2367         &  2.1468    &  1.4925     &  4.4335         &   0.7478   &   0.5140    &  1.0212         \\ 
FedDA ($\varphi$=$10$)                                   &  \textit{0.3481}    & 0.0753      &  0.1062         & \textit{0.3321}     & 0.1810      &  0.2275         &  \textit{2.0719}    &  \textit{1.1699}     &   \textit{3.9266} &  \textit{0.7320}    &   \textit{0.4543}    &  \textit{0.9504}         \\
FedDA ($\varphi$=$100$)                                   & \textbf{0.3322}     & \textbf{0.0659}      & \textbf{0.1033}          & \textbf{0.3214}     &      \textbf{0.1741} & \textbf{0.2211}          &  \textbf{1.9703}    & \textbf{1.0592}      &   \textbf{2.4473}        &  \textbf{0.6920}    &  \textbf{0.4281}     & \textbf{0.7471}          \\  \hline
\multicolumn{1}{|r|}{$\uparrow $ ($\varphi$=$100$)}                                       &   +9.4\%   &   +14.9\%     &   +5.8\%         & +4.8\%      &   +5.2\%     &   +4.7\%         &   +8.6\%    &   +33.7\%     &   +48.6\%         &   +7.0\%    &      +19.3\%  &   +29.7\%         \\ \hline
\end{tabular}%
}
\end{table*}

\begin{figure*}[!htbp]
\centering
\subfloat[Results on the Milano dataset for randomly selected cells.]
{\includegraphics[width=0.48\textwidth]{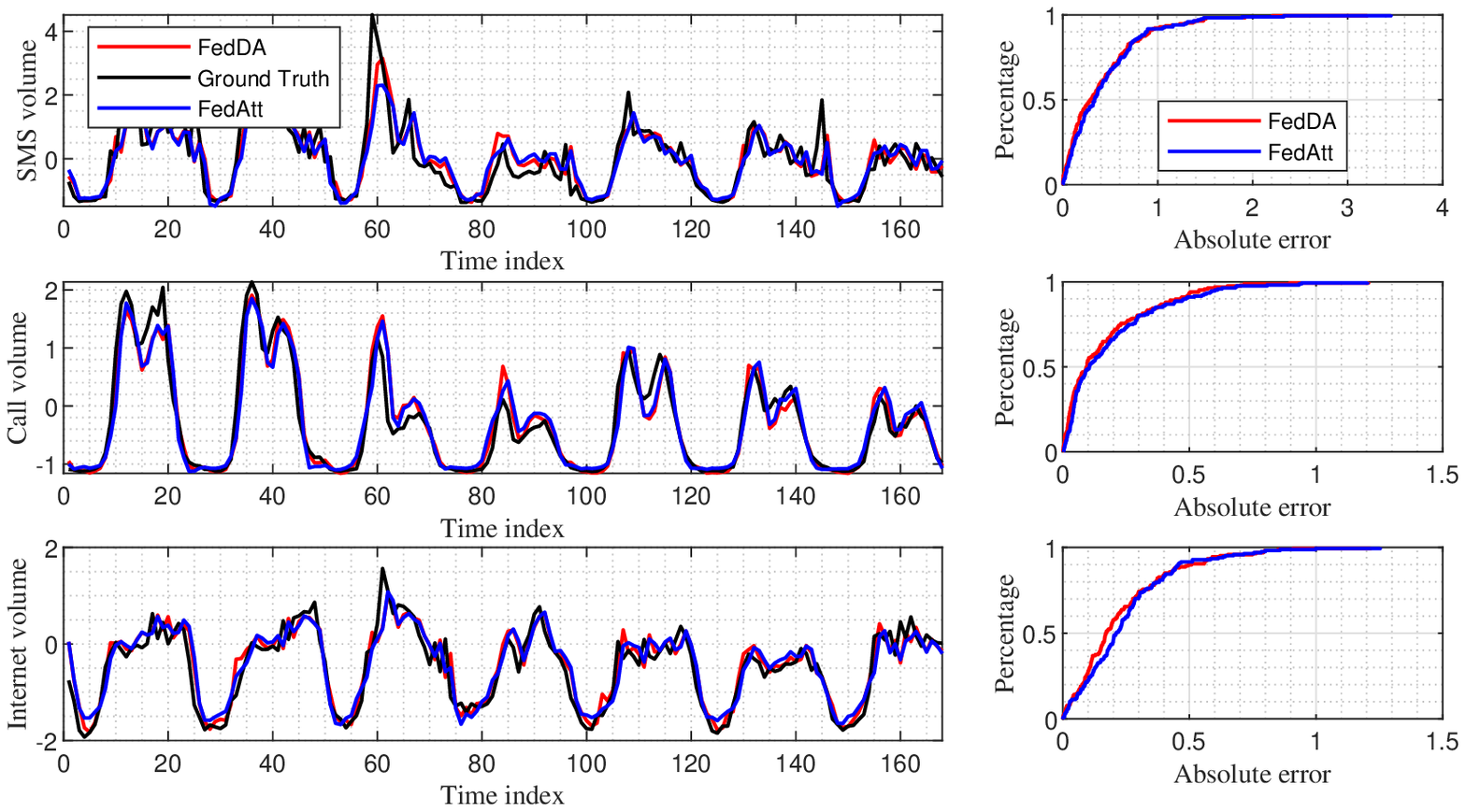}
\label{milano_curve}
}
\subfloat[Results on the Trento dataset for randomly selected cells.]
{\includegraphics[width=0.48\textwidth]{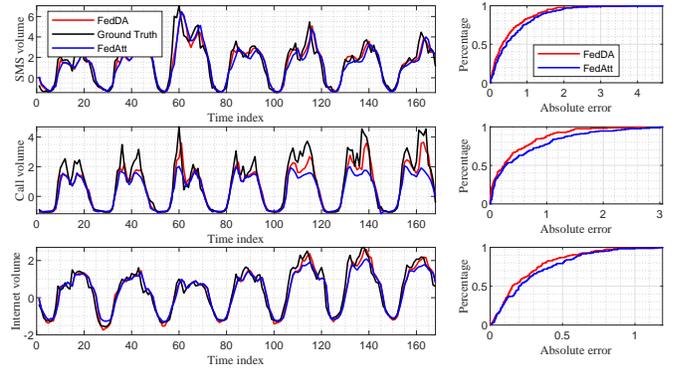}
\label{trento_curve}
}
\caption{Comparisons between predictions and the real values and the corresponding error analysis.}
\label{pred_curve}
\end{figure*}

\section{Experiments}
In this section, extensive experiments are conducted to validate the effectiveness and efficiency of FedDA. 
We begin with a brief introduction of the dataset, evaluation metrics, and baseline methods. Then, experimental settings, such as learning rate and batch size, are given. 
After that, we present experimental results, including the overall prediction performance of various methods and the influence of learning (hyper-)parameters on prediction performance.
\subsection{Dataset and Evaluation Metrics}
The datasets used in this paper come from the \textit{Big Data Challenge}\cite{Barlacchi2015} launched by Telecom Italia and mainly are Call Detail Records (CDR) of two Italian areas, i.e., the city of Milan and the province of Trentito.
The area of Milan is divided into a grid of $10000$ cells, and  for Trentino, the grid is of $6575$ cells.
In each cell, the user's telecommunication activities are served and logged by the BS and thus we use BS or cell interchangeably to denote a cell. There are threes types of wireless traffic, which are corresponding to SMS, voice Call, and Internet services. 
The traffic is logged every ten minutes over two months, from $11/01/2013$ to $01/01/2014$.
For experiments in the following subsections, the traffic is resampled into hourly to circumvent the data sparsity problem.
These two datasets are publicly available and can be accessed on Harvard Dataverse \cite{Italia2015a, Italia2015}.
To evaluate prediction performance, two widely used regression metrics are adopted in this paper, i.e., mean squared error (MSE) and mean absolute error (MAE). 
%

\subsection{Baseline Methods}
We compare our proposed wireless traffic prediction framework, FedDA, with five baseline methods.
\begin{itemize}
\item Lasso: A linear model for regression.
\item Support Vector Regression (SVR) \cite{Feng2006}: SVR is one of the most classical machine learning algorithms and has been successfully used for traffic prediction. 
\item LSTM \cite{Qiu2018}: LSTM has a strong ability to model time series dataset and normally has better prediction performance than linear models and shallow-learning models.
\item FedAvg \cite{McMahan2017}: FedAvg is first proposed in the pioneering work of federated learning. It adopts an average of local weights for model aggregation.
\item FedAtt \cite{Ji2019}: This algorithm is similar to FedAvg. However, when performing model aggregation in the central server, it differentiates and quantifies the contributions of different client models to the global model.
\end{itemize}
The first three baselines are trained in a fully distributed way. That is, the model is trained per client. The latter two baselines and our FedDA are trained in a federated way.

\subsection{Experimental Settings and Overall Results}
Without loss of generality, we randomly select $100$ cells from each dataset and carry out experiments on the three kinds of wireless traffic of these cells.
The traffic from the first seven weeks is used to train prediction models and the traffic from the last week is used for test.
When constructing training samples using sliding window scheme, the length of closeness dependence $p$ and periodicity dependence $q$ are both set to $3$.
Considering that the edge client has limited computing power and thermal constraints, a relatively lightweight LSTM network is adopted. Specifically, the network has two LSTM layers and each layer has $64$ hidden neurons, followed by a linear layer that maps the features to predictions.
All baselines except shallow learning algorithms share the same network architecture for the sake of fairness.
Unless otherwise specified, we take $100$ communication rounds between local clients and the central server and report results on the final model.
The regularization term $\rho$ is determined through a grid search with values ranging from $-0.3$ to $0.3$ and step size $0.1$.
The cluster size $C$ is set to $16$.
Similar to the standard settings in FL \cite{McMahan2017}, the values of local epochs and local batch size are set to $1$ and $20$, respectively. In each communication round, $10\%$ percent of the cells are involved in model training.
We use stochastic gradient descent (SGD) as the optimizer to update our models with learning rate $0.01$.

%

The experimental results of different prediction methods are presented in Table \ref{tab:performance}.
Note that in Table \ref{tab:performance}, our proposed method have three variations on the basis of how many augmented data samples shared, i.e., $\varphi$=$1$, $\varphi$=$10$, and $\varphi$=$100$. In reality, the amount of transfered data samples can be flexibly adjusted according to network situation.
It can be seen from this table that our proposed method, FedDA, outperforms all the baseline methods for all kinds of wireless traffic in both datasets, even with only $1\%$ of the augmentation data is shared. Here and in the following experiments, we report the results of FedDA with $\varphi$=$100$ unless otherwise specified. Specifically, for the SMS, CALL, and Internet service traffic of the Milano dataset, compared with the best-performing method in baselines, namely FedAtt, FedDA can offer MSE gains of $9.4\%$, $14.9\%$, and $5.8\%$, respectively.
Likewise, for the Trento dataset, FedDA yields performance gains of $8.6\%$ (SMS), $33.7\%$ (CALL), and $48.6\%$ (Internet), respectively.
In terms of the metric of MAE, though the improvements are not as remarkable as MSE, it is still obvious that an average of $4.9\%$ ($18.7\%$) performance gain can be achieved for the Milano (Trento) dataset. We can also notice that the prediction performance of FedDA improves consistently with the increase of shared augmentation data size. This is because the quasi-global model can capture the traffic patterns better when more data samples are available.
The success of FedDA can be attributed to the following reasons:
\begin{figure}[!htb]
\centering
\includegraphics[width=0.45\textwidth]{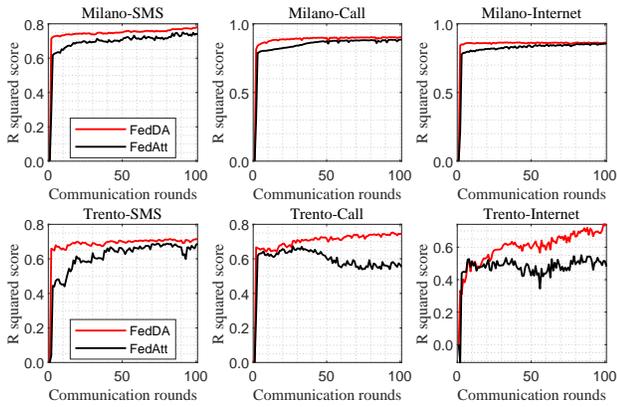}
\caption{Prediction accuracy versus communication rounds.}
\label{acc_curve}
\end{figure}
\begin{itemize}
    \item Compared with fully distributed algorithms (SVR and LSTM), which only consider temporal dependence of wireless traffic, FedDA can capture both the spatial dependence and the temporal dependence by means of model fusion, and is thus more robust;
    \item Compared with conventional FL algorithms (FedAvg and FedAtt), the introduced clustering strategy makes the learning process of FedDA case-specific, and the dual attention scheme greatly reduces data heterogeneity. Therefore, FedDA has a high generalization ability;
    \item FedDA can balance between capturing the unique characteristics of a cluster and the shared macro traffic patterns among different clusters, and thus has have more accurate predictions.
\end{itemize}

Besides, the shown results also indicate that in comparison with fully distributed algorithms, FL-based algorithms can achieve better predictions, and FedAtt achieves the second best prediction performance, followed by the classic FedAvg algorithm.
This is rather intuitive, since there is no knowledge sharing when training prediction models with fully distributed algorithms. The lack of knowledge sharing results in a loss of prediction accuracy.

%
%

To further evaluate the predictive ability of different algorithms, the comparisons between predicted values and the real values of different algorithms are given in Fig. \ref{pred_curve}.
The results on cumulative distribution functions (CDFs) of absolute prediction error are also included in Fig. \ref{pred_curve} for quantitatively measuring the goodness of prediction models.
Fig. \ref{milano_curve} (Fig. \ref{trento_curve}) represents results on the Milano (Trento) dataset.
More specifically, the left three subfigures of Fig. \ref{milano_curve} (Fig. \ref{trento_curve}) denote the comparisons between predictions and the ground truth for the SMS, Call, and Internet service traffic of randomly selected cells, and the right three subfigures are the corresponding CDFs of errors.
Here, we choose FedAtt as the benchmark for performance comparison, since it achieves the best performance among all baseline methods in Table \ref{tab:performance}.
By observing Fig. \ref{pred_curve}, we can tell that FedDA obtains consistent better prediction performance than FedAtt, on all three kinds of wireless traffic.
Meanwhile, it has smaller prediction errors, especially when the traffic volume comes to high and unstable.

For prediction errors, taking the SMS traffic of the Trento dataset for example, there are approximately $83\%$ errors that are smaller than $1$ for FedDA, while the case for FedAtt is about $76\%$. 
Moreover, the average prediction errors for FedAtt and FedDA are $0.65$ and $0.54$, respectively.
Based on the above evaluation, we can summarize that FedDA can achieve more accurate prediction results than those of baseline methods.

\begin{figure}[!htb]
\centering
\includegraphics[width=0.45\textwidth]{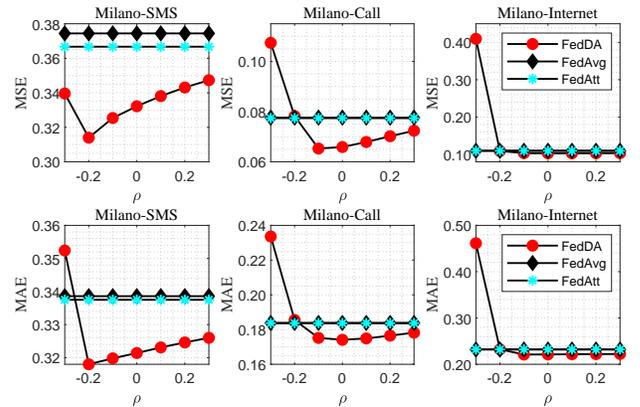}
\caption{Influence of $\rho$ on prediction performance for the Milano dataset. }
\label{milano_rho}
\end{figure}

\begin{figure}[!htb]
\centering
\includegraphics[width=0.45\textwidth]{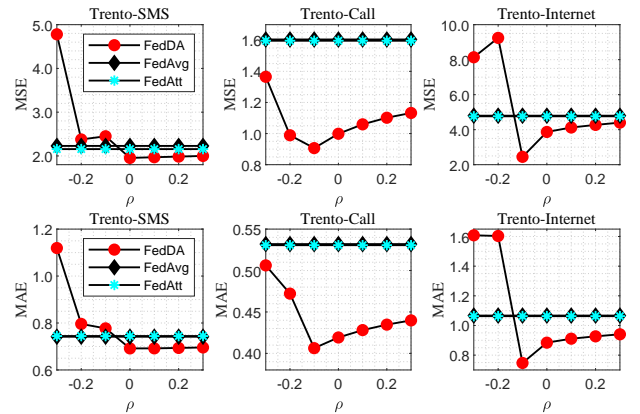}
\caption{Influence of $\rho$ on prediction performance for the Trento dataset.}
\label{trento_rho}
\end{figure}

\subsection{Communication Rounds Versus Prediction Accuracy}

\begin{figure*}[!htb]
\centering
\subfloat[][Results on the Milano dataset.]
{\includegraphics[width=0.35\textwidth]{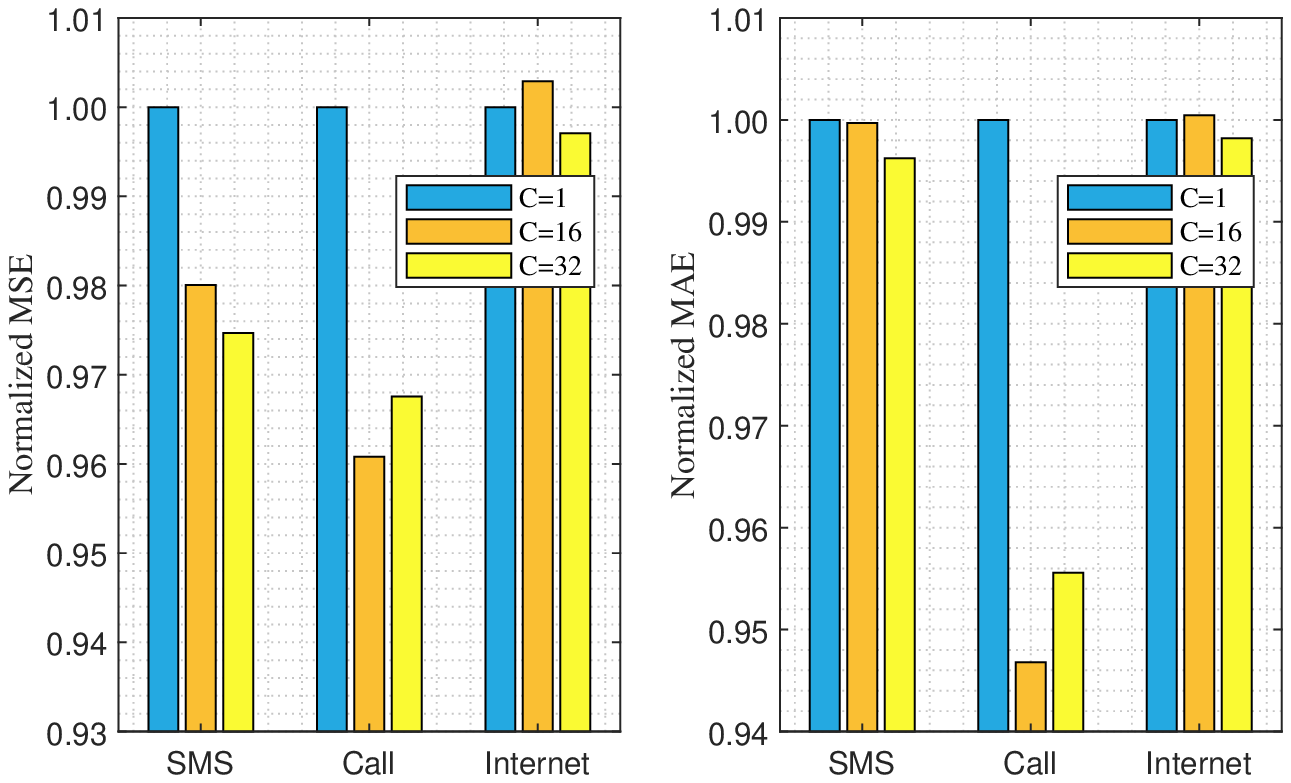}
\label{milano_cluster}
}
\subfloat[][Results on the Trento dataset.]
{\includegraphics[width=0.35\textwidth]{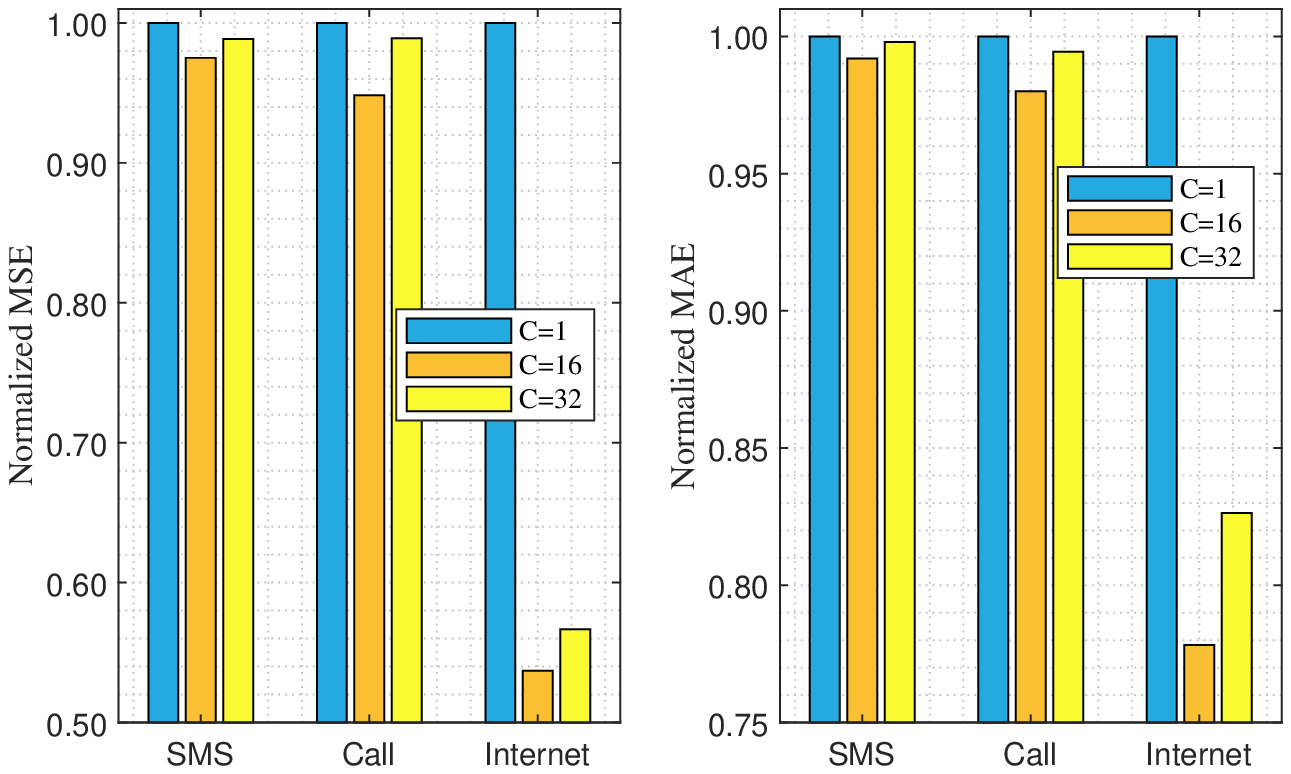}
\label{trento_cluster}
}
\caption{Cluster size versus prediction performance.}
\label{cluster_bar}
\end{figure*}

In FL or any other distributed learning frameworks, it is assumed that communication resources are more precious than computation resources and fewer communications are preferred. Thus, in this subsection, we report the prediction accuracy along with each communication round (epoch) and use R-squared score to denote the accuracy as it reflects how well ground truth values are predicted by the model \cite{Cameron1996}.
The obtained results are summarized in Fig. \ref{acc_curve}, in which the upper (lower) three subfigures represent results on the Milano (Trento) dataset.
From Fig. \ref{acc_curve} we can observe that FedDA achieves higher accuracy on both two datasets and its advantages are more clear on the Trento dataset. More importantly, FedDA needs much fewer communications to achieve a certain prediction accuracy. Take the Milano dataset as an example, after $30$ communication rounds, FedDA can achieve accuracies of $0.74$, $0.89$, and $0.86$ for the SMS, Call, and Internet traffic, respectively. While for FedAtt, the achieved accuracies for the SMS, Call, and Internet traffic are $0.69$, $0.84$, and $0.83$, respectively. Thus here we argue that our proposed method is communication-efficient, which is key for performing learning tasks at the edge.

\subsection{Effect of $\rho$ on Prediction Performance}
As $\rho$ is a task-dependent regularizer and it impacts the prediction performance a lot, we present the prediction performance of FedDA along with $\rho$ and summarize the results in Fig. \ref{milano_rho} and Fig. \ref{trento_rho} for the Milano and Trento datasets, respectively.
In both figures, the value of $\rho$ is ranging from $-0.3$ to $0.3$ with a step size of $0.1$ and the obtained MSE (MAE) results are given in the upper (lower) three subfigures. Besides FedDA, the other two FL-based methods, i.e., FedAvg and FedAtt, are also included for comparison purpose but their MSE and MAE results keep constant as they are not affected by $\rho$.
We can tell from Fig. \ref{milano_rho} and Fig. \ref{trento_rho} that for FedDA, with the increase of $\rho$, the values of MSE and MAE first decrease rapidly, and then slowly increase. For the other two baselines, FedAtt achieves slightly better results than FedAvg.
Though $\rho$ has a great influence on the prediction performance, FedDA generally can yield lower MSE and MAE values than FedAvg and FedAtt. 
Take the SMS traffic of the Milano dataset as an example, the obtained MSE results are always better than FedAvg and FedAtt regardless of the choice of $\rho$; while for the metric of MAE, similar conclusion hold except that $\rho$=$-0.3$, by which FedDA achieves worse results than FedAvg and FedAtt. Nonetheless, the optimal values of $\rho$ can be determined by using a grid search strategy during model training and the cost is low.
The results in these two figures demonstrate that the dual attention scheme in FedDA can indeed improve prediction performance by introducing prior knowledge and the influence varies on different datasets.

\subsection{Effect of $C$ on Prediction Performance}
The cluster size $C$ determines how many cells are involved in model aggregation and it also affects the final prediction performance.
Thus in this subsection, we explore how the cluster size affects the prediction performance of FedDA and the obtained MSE and MAE results are plotted in Fig. \ref{cluster_bar}.
In particlular, Fig. \ref{milano_cluster} (Fig. \ref{trento_cluster}) shows the results on the Milano (Trento) dataset.
We consider three scenarios, i.e., $C$=$1$, $C$=$16$, and $C$=$32$. Note that $C$=$1$ means no clustering adopted in prediction. In addition, to make the comparison clearer, the MSE and MAE results of $C$=$16$ and $C$=$32$ are normalized based on the results of $C$=$1$.
We can observe that the choices of $C$ yield different influences on the prediction performance of FedDA.
In most cases, introducing the clustering strategy can indeed lead to lower prediction errors.
Specifically, we can observe from Fig. \ref{milano_cluster} that FedDA achieves considerable performance improvements when cluster size is $16$ or $32$, for the SMS and Call traffic. For the Internet traffic, though the performance degrades slightly when $C$=$16$, it improves when $C$=$32$.
For the Trento dataset, introducing the clustering strategy can always yield better prediction performance than not, especially for the Internet traffic, on which the improvement is up to $50\%$.
Overall, the results in Fig. \ref{cluster_bar} demonstrate the superiority of introducing the clustering into FedDA.
This is because the cluster size $C$ controls the specialty of FedDA and thereby affects the global model. If no clustering strategy is involved, all data is mixed together to generate the global model. In this case, some unique traffic patterns hidden in the data cannot be captured by FedDA and hence leads to performance degradation.

\section{Conclusion}
In this work, we investigated the wireless traffic prediction problem and proposed a novel framework called FedDA. To deal with the heterogeneity of wireless traffic data, we proposed a data-sharing strategy in FedDA by transferring a small augmented traffic dataset to the central server, by which a quasi-global model is obtained and shared among all BSs. Besides, we also introduced an iterative clustering algorithm to cluster BSs into different groups, by considering both the wireless traffic pattern and the geo-location information. To enhance the generalization ability of the global model, we proposed a dual attention-based model aggregation scheme by paying attention to the unequal contributions of different local models and the quasi-global model. The aggregation scheme is applied over a hierarchical architecture so as to capture both intra-cluster and inter-cluster patterns of wireless data traffic. Finally, we verified the effectiveness and efficiency of FedDA on two real-world datasets.


{\small
\bibliography{fedda}
}

\end{document}